\documentclass[%
 reprint,
 superscriptaddress,
 nofootinbib,
 amsmath,amssymb,
 aps,
 prd,
 floatfix,
]{revtex4-2}

\usepackage{hyperref}
\usepackage{graphicx}
\usepackage{dcolumn}
\usepackage{bm}
\usepackage{color}

\begin{document}

\title{Effective Sub-Quantum Readout for Non-Monochromatic Axion Signals in High-$Q$ Haloscopes}

\author{Junu Jeong}
\email{jun-woo.jeoung@fysik.su.se}
\affiliation{The Oskar Klein Centre, Department of Physics, Stockholm University, AlbaNova, SE-10691 Stockholm, Sweden}

\author{Max~Silva-Feaver}
\affiliation{Department of Physics, Yale University, New Haven, Connecticut 06520, USA}
\affiliation{Wright Laboratory, Yale University, New Haven, Connecticut 06520, USA}

\date{\today}

\begin{abstract}
The search for wave-like dark matter using microwave cavity haloscopes is constrained by the Standard Quantum Limit, which dictates that phase-preserving linear amplification results in a minimum of one quantum of total system noise for a narrow-band signal.
We demonstrate that this limit is effectively halved for a non-monochromatic axion signal coupled to a high-$Q$ cavity.
By operating a Josephson Parametric Amplifier such that the cavity resonance is centered exactly at the half-pump frequency, the axion signal symmetrically populates both the signal and idler bands.
Through quadrature analysis of the homodyne readout, we show that the incoherent sum of these mirrored spectral components doubles the measured signal power while the vacuum noise remains constant.
This operation yields an effective noise limit of 0.5 quanta per frequency bin, translating to an overall effective limit of $1/\sqrt{2}$ quanta after optimal matched filtering.
\end{abstract}

\maketitle

\section{Introduction}

The axion is a hypothetical particle proposed to resolve the strong $CP$ problem in quantum chromodynamics~\cite{PecceiQuinn:PRL:1977,Weinberg:PRL:1978,Wilczek:PRL:1978} and is a strong candidate for cold dark matter~\cite{Preskill:PLB:1983,Abbott:PLB:1983,Dine:PLB:1983}.
The detection of axion dark matter via the Sikivie haloscope relies on measuring the microwave signals generated by axion-photon conversion inside a resonant cavity~\cite{Sikivie:PRL:1983}.
Because these signals are typically orders of magnitude weaker than standard electronic noise, modern experiments employ quantum-limited amplifiers, such as the Josephson Parametric Amplifier (JPA)~\cite{Yamamoto_jpa,lehnert_jpa}, to perform the initial readout~\cite{CAPP:PRX:2024,ADMX:PRL:2025,HAYSTAC:PRL:2025}. 

According to the theorem established by Caves~\cite{Caves1982}, a phase-preserving linear amplifier results in a total of at least one quantum of noise---half a quantum from the signal port vacuum fluctuations and half a quantum from the idler port.
However, this limit implicitly assumes a narrow-band (single-sideband) signal.
While experiments have utilized squeezed vacuum systems to operate below the standard quantum limit~\cite{Malnou2019,haystac_phaseIIa,HAYSTAC:PRL:2025}, these techniques mitigate the additional vacuum noise introduced by the readout chain (e.g., circulators) rather than the intrinsic quantum noise of the cavity itself.

In this Letter, we demonstrate that when the cavity quality factor significantly exceeds that of the axion ($Q_c \gg Q_a$)~\cite{Ahn:PRA:2022,QUAX:PRD:2022,Cervantes2022,ahn2023thesis}, the non-monochromatic nature of the axion signal can be exploited to effectively suppress this added noise.
By centering the cavity resonance exactly at the JPA half-pump frequency, the measured axion signal is doubled, just as the vacuum noise receives contributions from both ports.
This yields a factor-of-two enhancement in the signal-to-noise ratio for each frequency bin, corresponding to an effective system noise of 0.5 quanta when normalized to the signal power.
This enhancement for a broadband signal was demonstrated experimentally~\cite{Renger:npjQI:2021}; here, we establish the necessary conditions for its application in haloscopes within a formal mathematical framework.
Upon optimally combining neighboring frequency bins during data analysis, this enhancement translates into a twofold increase in the scanning rate.

\section{Quadratures of the JPA Output in a Homodyne Readout}
\label{sec:VXY}

Consider a JPA connected to an IQ mixer, where the local oscillator (LO) frequency is exactly equal to the half-pump frequency.
A similar approach applies to a heterodyne circuit, which requires summing the signal and idler components in the Fourier domain to effectively emulate homodyne detection~\cite{Eichler:PRL:2011,Esposito:PRL:2022}.
The output photons from the JPA are described by~\cite{Haus:PR:1962,Caves1982,Backes:Thesis:2021}:
\begin{equation}
\label{eq:jpa_out}
\begin{split}
    & \hat{a}_{\mathrm{out}}(\delta) = g_{s}(\delta) \hat{a}_{\mathrm{in}}(\delta) + g_{i}(\delta) \hat{a}_{\mathrm{in}}^{\dagger}(-\delta), \\
    & \hat{a}_{\mathrm{out}}^{\dagger}(-\delta) = g_{i}^{*}(-\delta) \hat{a}_{\mathrm{in}}(\delta) + g_{s}^{*}(-\delta) \hat{a}_{\mathrm{in}}^{\dagger}(-\delta),
\end{split}
\end{equation}
where $\delta$ is the detuning from the half-pump frequency, $f_{p}/2$, and the operators are evaluated in a frame rotating at $f_{p}/2$. Here, $g_{s}$ and $g_{i}$ are the signal and idler gains, which satisfy $\left| g_{s} (\delta) \right|^{2} - \left| g_{i} (\delta) \right|^{2} = 1$.
Furthermore, we define the degenerate gain parametrized by the detuning and an arbitrary phase $\theta$ (later identified with the phase of a specific input signal in Equation~\ref{eq:signal}) as:
\begin{equation}
    g_{d} (\delta, \theta) = g_{s}(\delta)e^{-i\theta} + g_{i}^{*}(-\delta) e^{i\theta}.
\end{equation}

If we consider a coherent input signal present solely at $f_{p}/2 + \delta$, the input at $-\delta$ is simply the vacuum state ($\langle \hat{a}_{\mathrm{in}}^{\dagger}(-\delta) \rangle = 0$). Even if an additional signal is present at $-\delta$, it would lack phase coherence with the primary signal at $+\delta$. Its interference contribution vanishes, allowing it to be treated independently. Assuming an input coherent amplitude of $\langle\hat{a}_{\mathrm{in}}(\delta)\rangle = A_0 e^{-i\theta}$ (where $A_{0}$ is real), the expectation values of the output operators defining the corresponding signal and idler envelope amplitudes, $A_{s}$ and $A_{i}$, are:
\begin{equation}
\label{eq:signal}
\begin{split}
    & \langle \hat{a}_{\mathrm{out}}(\delta) \rangle = g_{s}(\delta) A_{0} e^{-i\theta} \equiv A_{s}(\delta) e^{-i\theta},\\
    & \langle \hat{a}_{\mathrm{out}}^{\dagger}(-\delta) \rangle = g_{i}^{*}(-\delta) A_{0} e^{-i\theta} \equiv A_{i}(-\delta) e^{-i\theta}.
\end{split}
\end{equation}

For an IQ mixer where the LO is set exactly at the half-pump frequency, the output voltages consist of I and Q channels, rotated by an angle $\theta_{\mathrm{IQ}}$ to align with the JPA pump phase:
\begin{equation}
\begin{split}
    \label{eq:VXY}
    & V_{X}(\delta) = V_{I}(\delta)\cos\theta_{\mathrm{IQ}} - V_{Q}(\delta)\sin \theta_{\mathrm{IQ}}, \\
    & V_{Y}(\delta) = V_{I}(\delta)\sin\theta_{\mathrm{IQ}} + V_{Q}(\delta)\cos \theta_{\mathrm{IQ}}.
\end{split}
\end{equation}
Because the mixer in the homodyne setup folds the negative frequency (idler) and the positive frequency (signal) into the same baseband frequency $\delta$, the I and Q voltage components are:
\begin{equation}
    \label{eq:VIQ}
    \begin{split}
        V_{I} (\delta) &= \langle \hat{a}_{\mathrm{out}}(\delta)e^{-i\theta_{\mathrm{LO}}} + \hat{a}_{\mathrm{out}}^{\dagger}(-\delta)e^{i\theta_{\mathrm{LO}}} \rangle \\
        &= A_{s}(\delta)e^{-i(\theta+\theta_{\mathrm{LO}})} + A_{i}(-\delta)e^{-i(\theta-\theta_{\mathrm{LO}})}, \\
        V_{Q} (\delta) &= A_{s}(\delta)e^{-i(\theta+\theta_{\mathrm{LO}}+\pi/2)} + A_{i}(-\delta)e^{-i(\theta-\theta_{\mathrm{LO}}-\pi/2)}.
    \end{split}
\end{equation}
Substituting Equation~\ref{eq:VIQ} into Equation~\ref{eq:VXY} and defining the relative measurement phase $\Delta\theta = \theta_{\mathrm{LO}} - \theta_{\mathrm{IQ}}$, we can factor out the input signal phase $\theta$:
\begin{equation}
\begin{split}
    V_{X}(\delta) &= A_{0} e^{-i\theta} \left[ g_{s}(\delta) e^{-i\Delta\theta} + g_{i}^{*}(-\delta) e^{i\Delta\theta} \right] \\
    &= A_{0} e^{-i\theta} g_{d}(\delta, \Delta\theta), \\
    V_{Y}(\delta) &= A_{0} e^{-i\theta} g_{d}(\delta, \Delta\theta + \pi/2).
\end{split}
\end{equation}
This represents a single-quadrature measurement, where the effective gain is modulated by the measurement phase.

When we perform image rejection by combining the X and Y quadratures in the complex plane, we evaluate the positive and negative frequencies of the complex signal $V_{X+iY}$. Using the identities $V_X(-\delta) = V_X^*(\delta)$ and $V_Y(-\delta) = V_Y^*(\delta)$ for classical real voltages, we obtain:
\begin{equation}
    \begin{split}
        V_{X+iY}(\delta) &= \frac{1}{2}\left(V_{X}(\delta) + i V_{Y}(\delta) \right) \\
        &= A_{s}(\delta) e^{-i(\theta + \Delta\theta)} = A_{0} e^{-i(\theta + \Delta\theta)} g_{s}(\delta), \\
        V_{X+iY}(-\delta) &= \frac{1}{2} \left( V_{X}(-\delta) + i V_{Y}(-\delta) \right) \\
        &= \frac{1}{2} \left( V_{X}^{*}(\delta) + i V_{Y}^{*}(\delta) \right) \\
        &= A_{i}^{*}(-\delta) e^{i(\theta - \Delta\theta)} = A_{0} e^{i(\theta - \Delta\theta)} g_{i}(-\delta).
    \end{split}
\end{equation}
Because the signal phase is conserved in $V_{X+iY}(\delta)$ and the overall signal and idler gains are independent of the pump or LO phase, this digital processing effectively isolates the output upper sideband mode and recovers a phase-insensitive amplifier response for the selected sideband.
The detailed calculation of the signal-to-noise ratios for a single-sideband input is provided in Appendix~\ref{apdx:snr}.

\section{Effective Noise for a Finite-bandwidth Signal}

Let us now consider the case of a dark matter axion signal, which possesses a finite bandwidth determined by the axion spectral distribution $\mathcal{D}_{a}$ and the cavity response function $\mathcal{L}_{c}$. For a single measurement realization, the complex amplitude of the input field is~\cite{Kim:JHEP:2020}:
\begin{equation}
    \langle \hat{a}_{\rm in}(\delta) \rangle \sim \frac{e^{i\theta_{\delta}}}{\Delta\nu} \int_{\delta-\Delta\nu/2}^{\delta+\Delta\nu/2} \mathcal{D}_{a}(\nu_{a}, \delta^{\prime}) \mathcal{L}_{c}(\nu_{c}, \delta^{\prime}) d\delta^{\prime},
\end{equation}
where $\theta_{\delta}$ is the random phase of the dark matter field at frequency $\delta$, and $\Delta \nu$ is the resolution bandwidth. If the cavity quality factor is significantly higher than that of the dark matter axion ($Q_c \gg Q_a$), we can approximate $\mathcal{D}_{a}$ as roughly constant over the cavity resonance. The input field then simplifies to the Lorentzian response function when $\Delta\nu\ll f_{p}/(2 Q_{c})$:
\begin{equation}
    \langle \hat{a}_{\rm in}(\delta) \rangle \sim \mathcal{L}_{c}(\nu_{c}, \delta)e^{i\theta_{\delta}}.
\end{equation}

Under these conditions, the axion signal becomes symmetric around the half-pump frequency provided its mass falls near the cavity resonance. This symmetry implies that the signal components add incoherently in the same manner as the vacuum noise during standard power measurements. In standard homodyne detection, the Standard Quantum Limit (SQL) arises because the mixer folds mirrored frequency components into the baseband, doubling the quantum noise contribution. However, in this symmetrically aligned configuration, the axion signal populates both sidebands, meaning the measured signal power is doubled.
Because the effective signal power is doubled for the same output noise floor, the SNR improves by a factor of two---equivalently, the noise referred to the input signal is reduced to 0.5 quanta.
For low-$Q$ cavities, this condition is only satisfied in a narrow region around the half-pump frequency, rendering the overall improvement negligible over a wide cavity bandwidth.

To mathematically demonstrate the impact of the cavity detuning on the signal-to-noise ratio, we consider the Power Spectral Density (PSD) of the digitized signal after image rejection.
Because it is established that the ideal signal-to-noise ratio of a double-quadrature measurement equals that of a single-quadrature measurement~\cite{Malnou2019}, analyzing this complex signal provides a representation of the system's performance.
By elevating the classical voltages derived in Section~\ref{sec:VXY} to quantum operators, we observe that forming the image-rejected signal mathematically isolates the positive-frequency JPA output mode:
\begin{equation}
    \hat{V}_{X+iY}(\delta) = \frac{1}{2}\left( \hat{V}_{X}(\delta) + i \hat{V}_{Y}(\delta) \right) = \hat{a}_{\rm out}(\delta) e^{-i\Delta\theta}.
\end{equation}
Therefore, evaluating the PSD of the measured complex voltage is fundamentally equivalent to evaluating the PSD of the JPA output operator $\hat{a}_{\rm out}(\delta)$.

Let the input field operator be composed of the finite-linewidth axion signal and vacuum noise, $\hat{a}_{\rm in}(\delta) = \hat{s}(\delta) + \hat{v}(\delta)$, where the spectral density of the axion signal, $S_a(\delta)$, is defined by $\langle \hat{s}^{\dagger}(\delta)\hat{s}(\delta^{\prime}) \rangle = S_a(\delta)\delta(\delta - \delta^{\prime})$.

Standard spectrum analyzers measure the symmetrized PSD, defined as $S_{\rm out}(\delta)\delta(\delta - \delta^{\prime}) \equiv \frac{1}{2}\langle \{ \hat{a}_{\rm out}^\dagger(\delta), \hat{a}_{\rm out}(\delta^{\prime}) \} \rangle$. To evaluate this, we first expand the normally-ordered product using Equation~\ref{eq:jpa_out}:
\begin{equation}
\begin{split}
    \langle \hat{a}_{\rm out}^\dagger(\delta) \hat{a}_{\rm out}(\delta^{\prime}) \rangle &= g_s^*(\delta)g_s(\delta^{\prime}) \langle \hat{a}_{\rm in}^\dagger(\delta)\hat{a}_{\rm in}(\delta^{\prime}) \rangle \\
    &\quad + g_i^*(\delta)g_i(\delta^{\prime}) \langle \hat{a}_{\rm in}(-\delta) \hat{a}_{\rm in}^\dagger(-\delta^{\prime}) \rangle \\
    &\quad + g_s^*(\delta)g_i(\delta^{\prime}) \langle \hat{a}_{\rm in}^\dagger(\delta)\hat{a}_{\rm in}^\dagger(-\delta^{\prime}) \rangle \\
    &\quad + g_i^*(\delta)g_s(\delta^{\prime}) \langle \hat{a}_{\rm in}(-\delta)\hat{a}_{\rm in}(\delta^{\prime}) \rangle.
\end{split}
\end{equation}
Because the inputs lack correlation between the $+\delta$ and $-\delta$ bands, the cross-expectation values ($\langle \hat{a}^\dagger \hat{a}^\dagger \rangle$ and $\langle \hat{a} \hat{a} \rangle$) vanish. Substituting the input signal spectral densities and applying the Bosonic commutation relation for the vacuum $[\hat{v}(\delta), \hat{v}^{\dagger}(\delta^{\prime})] = \delta(\delta - \delta^{\prime})$, the normally-ordered output spectrum reduces to:
\begin{equation}
    \langle \hat{a}_{\rm out}^\dagger(\delta) \hat{a}_{\rm out}(\delta^{\prime}) \rangle = \Big( |g_s|^2 S_a(\delta) + |g_i|^2 [S_a(-\delta) + 1] \Big) \delta(\delta - \delta^{\prime}).
\end{equation}

By utilizing the output field commutation relation $[\hat{a}_{\rm out}(\delta), \hat{a}_{\rm out}^\dagger(\delta^{\prime})] = \delta(\delta - \delta^{\prime})$, the symmetrized PSD can be rewritten as:
\begin{equation}
\begin{split}
    S_{\rm out}(\delta) &= \frac{\langle \hat{a}_{\rm out}^\dagger(\delta) \hat{a}_{\rm out}(\delta^{\prime}) \rangle}{\delta(\delta - \delta^{\prime})} + \frac{1}{2} \\
    &= |g_s|^2 S_a(\delta) + |g_i|^2 S_a(-\delta) + |g_i|^2 + \frac{1}{2}.
\end{split}
\end{equation}
By rewriting the final constant term as $\frac{1}{2} = \frac{1}{2}(|g_s|^2 - |g_i|^2)$, the equation groups into a symmetric form:
\begin{equation}
    S_{\rm out}(\delta) = |g_s|^2 \left[ S_a(\delta) + \frac{1}{2} \right] + |g_i|^2 \left[ S_a(-\delta) + \frac{1}{2} \right].
\end{equation}
In this symmetric form, the two $1/2$ terms carry distinct physical origins in accordance with the standard quantum limit.
The $1/2$ in the first term represents the intrinsic input zero-point fluctuations of the signal mode, which exist independent of amplification.
The $1/2$ in the second term originates from the vacuum fluctuations entering the idler port, representing the added noise that the phase-preserving amplifier injects.

To establish a baseline, we first consider the standard non-degenerate case where the cavity resonance is detuned from the half-pump frequency ($\nu_c \neq 0$). In this regime, the cavity filters the axion signal such that it occupies only a single sideband (e.g., $S_a(\delta) \equiv S_a$ while $S_a(-\delta) \approx 0$). In the high-gain limit where $|g_s|^2 \approx |g_i|^2 \equiv G$, the output PSD evaluates to:
\begin{equation}
    S_{\rm out}^{\nu_c \neq 0}(\delta) \approx G \left[ S_a + \frac{1}{2} \right] + G \left[ 0 + \frac{1}{2} \right] = G (S_a + 1).
\end{equation}
Referred back to the input signal power $S_a$, the noise power is equivalent to $1$ quantum. This recovers the SQL dictated by Caves for a standard narrow-band, single-sideband signal~\cite{Caves1982}.

In contrast, in the quasi-degenerate scenario where the cavity resonance is centered at the half-pump frequency ($\nu_{c} = 0$), the non-monochromatic axion signal populates both the upper ($+\delta$) and lower ($-\delta$) sidebands symmetrically, such that $S_a(\delta) \approx S_a(-\delta) \equiv S_a$. The output PSD then simplifies to:
\begin{equation}
    S_{\rm out}^{\nu_c = 0}(\delta) \approx G \left[ S_a + \frac{1}{2} \right] + G \left[ S_a + \frac{1}{2} \right] = G (2 S_a + 1).
\end{equation}
The first term ($2G S_a$) represents the total amplified axion signal, incoherently gathered from both sidebands. The second term ($G$) remains the standard vacuum noise power floor. By referring the signal-to-noise ratio back to the input signal power $S_a$, the signal has doubled relative to the vacuum noise. Thus, the effective noise limit is halved to 0.5 quanta. 

An optimal matched filter is subsequently applied to the acquired spectrum to maximize the integrated signal-to-noise ratio. In the non-degenerate configuration, the filter integrates over the full Lorentzian lineshape. Conversely, in the quasi-degenerate configuration, the homodyne folding merges the symmetric halves of the Lorentzian, meaning the filter integrates over half the number of independent frequency bins. While integrating over twice as many independent data points provides the non-degenerate case with a relative statistical enhancement of $\sqrt{2}$ in SNR, the quasi-degenerate configuration intrinsically doubles the signal-to-noise ratio per bin. Consequently, the final integrated SNR of the quasi-degenerate setup remains larger by an overall factor of $\sqrt{2}$. Since the experimental scanning rate scales with the square of the SNR, this configuration increases the search rate by a factor of two.

\section{Conclusion}

For a standard phase-preserving measurement of a narrow-band signal, the minimum system noise is one quantum.
However, in the high-$Q$ quasi-degenerate scenario, the axion signal is symmetrically distributed across both sidebands.
Because the signal components add incoherently in the same manner as the vacuum noise, the measured signal power is doubled compared to a single-sideband input while the vacuum noise floor remains unchanged.
The signal-to-noise ratio per frequency bin is therefore enhanced by a factor of two, yielding an effective quantum limit of 0.5 quanta for a symmetrically distributed signal.
Because the homodyne folding halves the number of independent frequency bins, applying an optimal matched filter across the spectral lineshape yields an overall integrated signal-to-noise ratio enhancement of $\sqrt{2}$.
This translates into a twofold increase in the scanning rate for high-$Q$ wave-like dark matter searches~\cite{QUAX:PRD:2022,Cervantes2022,ahn2023thesis}, effectively overcoming the narrow-band SQL.

\begin{acknowledgments}
This research is supported by the Knut and Alice Wallenberg Foundation.
The authors gratefully acknowledge valuable discussions within the ALPHA Collaboration.
\end{acknowledgments}

\appendix

\section{Signal-to-Noise Ratio in the Presence of Post-Amplifier Noise}
\label{apdx:snr}

To quantify the robustness of this readout scheme against downstream losses, let us evaluate the signal-to-noise ratio (SNR) in the presence of a secondary amplifier, such as a high-electron-mobility transistor (HEMT).
We define $S_0 = A_0^2$ as the input signal power and $N_{\mathrm{vac}} = 1/2$ as the fundamental vacuum noise variance (half a quantum).
In the high-gain limit of the JPA, we assume $|g_s(\delta)|^2 \approx |g_i(\delta)|^2 \approx G$.
The HEMT amplifier adds independent, uncorrelated classical thermal noise to both the $+\delta$ and $-\delta$ sidebands.
Let this added noise variance be $N_{\mathrm{HEMT}}$ per sideband.

Prior to the homodyne mixer, the signal sideband ($+\delta$) contains the amplified signal ($G S_0$), the amplified vacuum noise ($G (N_{\mathrm{vac}}(\delta) + N_{\mathrm{vac}}(-\delta))$), and the HEMT noise ($N_{\mathrm{HEMT}}$).
The idler sideband ($-\delta$) also contains the correlated idler signal ($G S_0$), the amplified vacuum noise ($G (N_{\mathrm{vac}}(\delta) + N_{\mathrm{vac}}(-\delta))$), and independent HEMT noise ($N_{\mathrm{HEMT}}$). The total power in each sideband is therefore:
\begin{equation}
\label{eq:total_p}
\begin{split}
    & P(+\delta) = G S_{0} + G (N_{\mathrm{vac}}(\delta) + N_{\mathrm{vac}}(-\delta)) + N_{\mathrm{HEMT}}(\delta), \\
    & P(-\delta) = G S_{0} + G (N_{\mathrm{vac}}(\delta) + N_{\mathrm{vac}}(-\delta)) + N_{\mathrm{HEMT}}(-\delta).
\end{split}
\end{equation}

In the single-quadrature (1Q) measurement, we isolate the $V_X(\delta)$ projection.
As derived in Section~\ref{sec:VXY}, when the measurement phase is optimally aligned ($\Delta\theta = 0$), the phase-correlated signal and idler components add coherently in voltage.
The resulting signal power is:
\begin{equation}
    P_{S,\mathrm{1Q}} \propto \left( \sqrt{G S_0} + \sqrt{G S_0} \right)^2 = 4G S_0.
\end{equation}
The quantum vacuum noise added by the JPA originates from the two-mode squeezing process, meaning that the vacuum fluctuations at $+\delta$ and $-\delta$ are correlated.
Along the amplified quadrature, this noise also adds coherently, effectively quadrupling the variance to $4G (N_{\mathrm{vac}}(\delta) + N_{\mathrm{vac}}(-\delta))$. 
This coherent homodyne folding doubles the effective noise bandwidth, yielding the standard quantum limit.
However, the HEMT noise components at $+\delta$ and $-\delta$ are uncorrelated.
When folded into the baseband by the mixer, their variances add incoherently:
\begin{equation}
    P_{N,\mathrm{1Q}}^{\mathrm{HEMT}} \propto N_{\mathrm{HEMT}}(\delta) + N_{\mathrm{HEMT}}(-\delta) = 2N_{\mathrm{HEMT}}.
\end{equation}
Thus, the effective SNR for the 1Q measurement is:
\begin{equation}
\begin{split}
    \mathrm{SNR}_{\mathrm{1Q}} &= \frac{4G S_0}{4G (N_{\mathrm{vac}}(\delta) + N_{\mathrm{vac}}(-\delta) ) + 2N_{\mathrm{HEMT}}} \\
    &= \frac{S_0}{2N_{\mathrm{vac}} + \frac{N_{\mathrm{HEMT}}}{2G}}.
\end{split}
\end{equation}

Conversely, for the double-quadrature (2Q) measurement, we perform image rejection by evaluating $V_{X+iY}(\delta)$ to isolate the positive-frequency mode.
This enables the direct application of Equation~\ref{eq:total_p}.
As demonstrated in Section~\ref{sec:VXY}, the signal power is proportional to $|A_0 g_s(\delta)|^2 \approx G S_0$.
The vacuum noise is evaluated similarly; however, summing the contributions from the signal and idler sidebands yields a total vacuum noise power of $2G N_{\mathrm{vac}}$.
Since the HEMT noise components in two quadratures are statistically independent (each carrying a variance of $N_{\mathrm{HEMT}}/2$), combining them results in the sum of their variances ($N_{\mathrm{HEMT}}$).
The effective SNR for the 2Q measurement is therefore:
\begin{equation}
    \mathrm{SNR}_{\mathrm{2Q}} = \frac{G S_0}{2G N_{\mathrm{vac}} + N_{\mathrm{HEMT}}} = \frac{S_0}{2N_{\mathrm{vac}} + \frac{N_{\mathrm{HEMT}}}{G}}.
\end{equation}

While the negative frequency component $V_{X+iY}(-\delta)$ carries identical signal information, its associated HEMT noise originates from the opposite sideband and is statistically independent. 
Averaging the power spectra of the positive and negative frequencies reduces the statistical fluctuation of the HEMT noise estimate by a factor of $\sqrt{2}$, but this power averaging does not recover the intrinsic SNR advantage of the 1Q measurement. 
However, if one averages the measurements at the complex voltage level, the sum $V_{X+iY}(\delta) + V_{X+iY}^{*}(-\delta)$ analytically reduces back to the real projection $V_X(\delta)$, naturally recovering the 1Q measurement and its optimal signal-to-noise ratio.

Comparing the two results, both methods reach the same fundamental quantum noise limit ($2 N_{\mathrm{vac}}$, equivalent to 1 quantum)~\cite{Malnou2019}.
However, the 1Q projection method leverages the phase coherence of the JPA output to quadruple the signal power ($4\times$) while only doubling the uncorrelated HEMT noise power ($2\times$).
As a result, a 1Q measurement suppresses the effective HEMT noise contribution by a factor of two relative to a 2Q measurement.
If the JPA gain $G$ is sufficiently high, the HEMT noise term $N_{\mathrm{HEMT}}/G$ vanishes, rendering the distinction between the two processing methods negligible.

\bibliography{main}

@article{PecceiQuinn:PRL:1977,
  title = {$\mathrm{CP}$ Conservation in the Presence of Pseudoparticles},
  author = {Peccei, R. D. and Quinn, Helen R.},
  journal = {Phys. Rev. Lett.},
  volume = {38},
  issue = {25},
  pages = {1440--1443},
  numpages = {0},
  year = {1977},
  month = {Jun},
  publisher = {American Physical Society},
  doi = {10.1103/PhysRevLett.38.1440},
  url = {https://link.aps.org/doi/10.1103/PhysRevLett.38.1440}
}

@article{Weinberg:PRL:1978,
  title = {A New Light Boson?},
  author = {Weinberg, Steven},
  journal = {Phys. Rev. Lett.},
  volume = {40},
  issue = {4},
  pages = {223--226},
  numpages = {0},
  year = {1978},
  month = {Jan},
  publisher = {American Physical Society},
  doi = {10.1103/PhysRevLett.40.223},
  url = {https://link.aps.org/doi/10.1103/PhysRevLett.40.223}
}

@article{Wilczek:PRL:1978,
  title = {Problem of Strong $P$ and $T$ Invariance in the Presence of Instantons},
  author = {Wilczek, F.},
  journal = {Phys. Rev. Lett.},
  volume = {40},
  issue = {5},
  pages = {279--282},
  numpages = {0},
  year = {1978},
  month = {Jan},
  publisher = {American Physical Society},
  doi = {10.1103/PhysRevLett.40.279},
  url = {https://link.aps.org/doi/10.1103/PhysRevLett.40.279}
}

@article{Preskill:PLB:1983,
title = {Cosmology of the invisible axion},
journal = {Phys. Lett. B},
volume = {120},
number = {1},
pages = {127-132},
year = {1983},
issn = {0370-2693},
doi = {https://doi.org/10.1016/0370-2693(83)90637-8},
url = {https://www.sciencedirect.com/science/article/pii/0370269383906378},
author = {John Preskill and Mark B. Wise and Frank Wilczek}
}

@article{Abbott:PLB:1983,
title = {A cosmological bound on the invisible axion},
journal = {Phys. Lett. B},
volume = {120},
number = {1},
pages = {133-136},
year = {1983},
issn = {0370-2693},
doi = {https://doi.org/10.1016/0370-2693(83)90638-X},
url = {https://www.sciencedirect.com/science/article/pii/037026938390638X},
author = {L.F. Abbott and P. Sikivie}
}

@article{Dine:PLB:1983,
title = {The not-so-harmless axion},
journal = {Phys. Lett. B},
volume = {120},
number = {1},
pages = {137-141},
year = {1983},
issn = {0370-2693},
doi = {https://doi.org/10.1016/0370-2693(83)90639-1},
url = {https://www.sciencedirect.com/science/article/pii/0370269383906391},
author = {Michael Dine and Willy Fischler}
}

@article{Sikivie:PRL:1983,
  title = {Experimental Tests of the "Invisible" Axion},
  author = {Sikivie, P.},
  journal = {Phys. Rev. Lett.},
  volume = {51},
  issue = {16},
  pages = {1415--1417},
  numpages = {0},
  year = {1983},
  month = {Oct},
  publisher = {American Physical Society},
  doi = {10.1103/PhysRevLett.51.1415},
  url = {https://link.aps.org/doi/10.1103/PhysRevLett.51.1415}
}

@ARTICLE{lehnert_jpa,
       author = {{Castellanos-Beltran}, M.~A. and {Lehnert}, K.~W.},
        title = "{Widely tunable parametric amplifier based on a superconducting quantum interference device array resonator}",
      journal = {Appl. Phys. Lett.},
         year = 2007,
        month = aug,
       volume = {91},
       number = {8},
          eid = {083509},
        pages = {083509},
          doi = {10.1063/1.2773988},
archivePrefix = {arXiv},
       eprint = {0706.2373},
 primaryClass = {cond-mat.supr-con},
       adsurl = {https://ui.adsabs.harvard.edu/abs/2007ApPhL..91h3509C}
}

@article{Yamamoto_jpa,
    author = {Yamamoto, T. and Inomata, K. and Watanabe, M. and Matsuba, K. and Miyazaki, T. and Oliver, W. D. and Nakamura, Y. and Tsai, J. S.},
    title = {Flux-driven Josephson parametric amplifier},
    journal = {Appl. Phys. Lett.},
    volume = {93},
    number = {4},
    pages = {042510},
    year = {2008},
    month = {07},
    issn = {0003-6951},
    doi = {10.1063/1.2964182},
    url = {https://doi.org/10.1063/1.2964182},
}

@article{CAPP:PRX:2024,
  title = {Extensive Search for Axion Dark Matter over 1 GHz with CAPP'S Main Axion Experiment},
  author = {Ahn, Saebyeok and Kim, JinMyeong and Ivanov, Boris I. and Kwon, Ohjoon and Byun, HeeSu and van Loo, Arjan F. and Park, SeongTae and Jeong, Junu and Lee, Soohyung and Kim, Jinsu and Kutlu, \ifmmode \mbox{\c{C}}\else \c{C}\fi{}a\ifmmode \breve{g}\else \u{g}\fi{}lar and Yi, Andrew K. and Nakamura, Yasunobu and Oh, Seonjeong and Ahn, Danho and Bae, SungJae and Choi, Hyoungsoon and Choi, Jihoon and Chong, Yonuk and Chung, Woohyun and Gkika, Violeta and Kim, Jihn E. and Kim, Younggeun and Ko, Byeong Rok and Miceli, Lino and Lee, Doyu and Lee, Jiwon and Lee, Ki Woong and Lee, MyeongJae and Matlashov, Andrei and Parashar, Pallavi and Seong, Taehyeon and Shin, Yun Chang and Uchaikin, Sergey V. and Youn, SungWoo and Semertzidis, Yannis K.},
  journal = {Phys. Rev. X},
  volume = {14},
  issue = {3},
  pages = {031023},
  numpages = {32},
  year = {2024},
  month = {Aug},
  publisher = {American Physical Society},
  doi = {10.1103/PhysRevX.14.031023},
  url = {https://link.aps.org/doi/10.1103/PhysRevX.14.031023}
}

@article{ADMX:PRL:2025,
  title = {{ADMX Axion Dark Matter Bounds around $3.3\text{ }\text{ }\mathrm{\ensuremath{\mu}}\mathrm{eV}$ with Dine-Fischler-Srednicki-Zhitnitsky Discovery Ability}},
  author = {Goodman, C. and Guzzetti, M. and Hanretty, C. and Rosenberg, L. J. and Rybka, G. and Sinnis, J. and Zhang, D. and Clarke, John and Siddiqi, I. and Chou, A. S. and Hollister, M. and Knirck, S. and Sonnenschein, A. and Caligiure, T. J. and Gleason, J. R. and Hipp, A. T. and Sikivie, P. and Solano, M. E. and Sullivan, N. S. and Tanner, D. B. and Khatiwada, R. and Carosi, G. and Cisneros, C. and Du, N. and Robertson, N. and Woollett, N. and Duffy, L. D. and Boutan, C. and Braine, T. and Lentz, E. and Oblath, N. S. and Taubman, M. S. and Daw, E. J. and Mostyn, C. and Perry, M. G. and Bartram, C. and Dyson, T. A. and Ruppert, S. and Withers, M. O. and Kuo, C. L. and McAllister, B. T. and Buckley, J. H. and Gaikwad, C. and Hoffman, J. and Murch, K. and Goryachev, M. and Hartman, E. and Quiskamp, A. and Tobar, M. E.},
  collaboration = {ADMX Collaboration},
  journal = {Phys. Rev. Lett.},
  volume = {134},
  issue = {11},
  pages = {111002},
  numpages = {7},
  year = {2025},
  month = {Mar},
  publisher = {American Physical Society},
  doi = {10.1103/PhysRevLett.134.111002},
  url = {https://link.aps.org/doi/10.1103/PhysRevLett.134.111002}
}

@article{HAYSTAC:PRL:2025,
  title = {{Dark Matter Axion Search with HAYSTAC Phase II}},
  author = {Bai, Xiran and Jewell, M. J. and Echevers, J. and van Bibber, K. and Droster, A. and Esmat, Maryam H. and Ghosh, Sumita and Graham, Eleanor and Jackson, H. and Laffan, Claire and Lamoreaux, S. K. and Leder, A. F. and Lehnert, K. W. and Lewis, S. M. and Maruyama, R. H. and Nath, R. D. and Rapidis, N. M. and Ruddy, E. P. and Silva-Feaver, M. and Simanovskaia, M. and Singh, Sukhman and Speller, D. H. and Zacarias, Sabrina and Zhu, Yuqi},
  collaboration = {HAYSTAC Collaboration},
  journal = {Phys. Rev. Lett.},
  volume = {134},
  issue = {15},
  pages = {151006},
  numpages = {8},
  year = {2025},
  month = {Apr},
  publisher = {American Physical Society},
  doi = {10.1103/PhysRevLett.134.151006},
  url = {https://link.aps.org/doi/10.1103/PhysRevLett.134.151006}
}

@article{Caves1982,
  title = {Quantum limits on noise in linear amplifiers},
  author = {Caves, Carlton M.},
  journal = {Phys. Rev. D},
  volume = {26},
  issue = {8},
  pages = {1817--1839},
  numpages = {0},
  year = {1982},
  month = {Oct},
  publisher = {American Physical Society},
  doi = {10.1103/PhysRevD.26.1817},
  url = {https://link.aps.org/doi/10.1103/PhysRevD.26.1817}
}

@article{Haus:PR:1962,
  title = {Quantum Noise in Linear Amplifiers},
  author = {Haus, H. A. and Mullen, J. A.},
  journal = {Phys. Rev.},
  volume = {128},
  issue = {5},
  pages = {2407--2413},
  numpages = {0},
  year = {1962},
  month = {Dec},
  publisher = {American Physical Society},
  doi = {10.1103/PhysRev.128.2407},
  url = {https://link.aps.org/doi/10.1103/PhysRev.128.2407}
}

@article{Malnou2019,
  title = {Squeezed Vacuum Used to Accelerate the Search for a Weak Classical Signal},
  author = {Malnou, M. and Palken, D. A. and Brubaker, B. M. and Vale, Leila R. and Hilton, Gene C. and Lehnert, K. W.},
  journal = {Phys. Rev. X},
  volume = {9},
  issue = {2},
  pages = {021023},
  numpages = {17},
  year = {2019},
  month = {May},
  publisher = {American Physical Society},
  doi = {10.1103/PhysRevX.9.021023},
  url = {https://link.aps.org/doi/10.1103/PhysRevX.9.021023}
}

@article{haystac_phaseIIa,
    author = "Backes, K. M. and others",
    collaboration = "HAYSTAC",
    title = "{A quantum-enhanced search for dark matter axions}",
    eprint = "2008.01853",
    archivePrefix = "arXiv",
    primaryClass = "quant-ph",
    doi = "10.1038/s41586-021-03226-7",
    journal = "Nature",
    volume = "590",
    number = "7845",
    pages = "238--242",
    year = "2021"
}

@phdthesis{Backes:Thesis:2021,
  author       = {Backes, Kelly Marie},
  title        = {A quantum-enhanced search for dark matter axions},
  school       = {Yale University},
  year         = {2021},
  month        = {October},
  url          = {https://elischolar.library.yale.edu/gsas_dissertations/294},
}

@article{Kim:JHEP:2020,
doi = {10.1088/1475-7516/2020/03/066},
url = {https://doi.org/10.1088/1475-7516/2020/03/066},
year = {2020},
month = {mar},
publisher = {},
volume = {2020},
number = {03},
pages = {066},
author = {Kim, Dongok and Jeong, Junu and Youn, SungWoo and Kim, Younggeun and Semertzidis, Yannis K.},
title = {Revisiting the detection rate for axion haloscopes},
journal = {J. Cosmol. Astropart. Phys.},
}

@article{Ahn:PRA:2022,
  title = {{Biaxially Textured ${\mathrm{YBa}}_{2}{\mathrm{Cu}}_{3}{\mathrm{O}}_{7\ensuremath{-}x}$ Microwave Cavity in a High Magnetic Field for a Dark-Matter Axion Search}},
  author = {Ahn, Danho and Kwon, Ohjoon and Chung, Woohyun and Jang, Wonjun and Lee, Doyu and Lee, Jhinhwan and Youn, Sung Woo and Byun, HeeSu and Youm, Dojun and Semertzidis, Yannis K.},
  journal = {Phys. Rev. Appl.},
  volume = {17},
  issue = {6},
  pages = {L061005},
  numpages = {6},
  year = {2022},
  month = {Jun},
  publisher = {American Physical Society},
  doi = {10.1103/PhysRevApplied.17.L061005},
  url = {https://link.aps.org/doi/10.1103/PhysRevApplied.17.L061005}
}

@article{QUAX:PRD:2022,
  title = {Search for Galactic axions with a high-$Q$ dielectric cavity},
  author = {Alesini, D. and Babusci, D. and Braggio, C. and Carugno, G. and Crescini, N. and D'Agostino, D. and D'Elia, A. and Di Gioacchino, D. and Di Vora, R. and Falferi, P. and Gambardella, U. and Gatti, C. and Iannone, G. and Ligi, C. and Lombardi, A. and Maccarrone, G. and Ortolan, A. and Pengo, R. and Rettaroli, A. and Ruoso, G. and Taffarello, L. and Tocci, S.},
  journal = {Phys. Rev. D},
  volume = {106},
  issue = {5},
  pages = {052007},
  numpages = {8},
  year = {2022},
  month = {Sep},
  publisher = {American Physical Society},
  doi = {10.1103/PhysRevD.106.052007},
  url = {https://link.aps.org/doi/10.1103/PhysRevD.106.052007}
}

@article{Cervantes2022,
  title = {Deepest sensitivity to wavelike dark photon dark matter with superconducting radio frequency cavities},
  author = {Cervantes, R. and Aumentado, J. and Braggio, C. and Giaccone, B. and Frolov, D. and Grassellino, A. and Harnik, R. and Lecocq, F. and Melnychuk, O. and Pilipenko, R. and Posen, S. and Romanenko, A.},
  journal = {Phys. Rev. D},
  volume = {110},
  issue = {4},
  pages = {043022},
  numpages = {10},
  year = {2024},
  month = {Aug},
  publisher = {American Physical Society},
  doi = {10.1103/PhysRevD.110.043022},
  url = {https://link.aps.org/doi/10.1103/PhysRevD.110.043022}
}

@phdthesis{ahn2023thesis,
  author       = {Ahn, Danho},
  title        = {(The) first high-temperature superconducting cavities in axion dark matter search},
  school       = {Korea Advanced Institute of Science and Technology},
  year         = {2023},
  url          = {http://hdl.handle.net/10203/307995}
}

@article{Eichler:PRL:2011,
  title = {Observation of Two-Mode Squeezing in the Microwave Frequency Domain},
  author = {Eichler, C. and Bozyigit, D. and Lang, C. and Baur, M. and Steffen, L. and Fink, J. M. and Filipp, S. and Wallraff, A.},
  journal = {Phys. Rev. Lett.},
  volume = {107},
  issue = {11},
  pages = {113601},
  numpages = {5},
  year = {2011},
  month = {Sep},
  publisher = {American Physical Society},
  doi = {10.1103/PhysRevLett.107.113601},
  url = {https://link.aps.org/doi/10.1103/PhysRevLett.107.113601}
}

@article{Esposito:PRL:2022,
  title = {Observation of Two-Mode Squeezing in a Traveling Wave Parametric Amplifier},
  author = {Esposito, Martina and Ranadive, Arpit and Planat, Luca and Leger, S\'ebastien and Fraudet, Dorian and Jouanny, Vincent and Buisson, Olivier and Guichard, Wiebke and Naud, C\'ecile and Aumentado, Jos\'e and Lecocq, Florent and Roch, Nicolas},
  journal = {Phys. Rev. Lett.},
  volume = {128},
  issue = {15},
  pages = {153603},
  numpages = {7},
  year = {2022},
  month = {Apr},
  publisher = {American Physical Society},
  doi = {10.1103/PhysRevLett.128.153603},
  url = {https://link.aps.org/doi/10.1103/PhysRevLett.128.153603}
}

@article{Renger:npjQI:2021,
author={Renger, M.
and Pogorzalek, S.
and Chen, Q.
and Nojiri, Y.
and Inomata, K.
and Nakamura, Y.
and Partanen, M.
and Marx, A.
and Gross, R.
and Deppe, F.
and Fedorov, K. G.},
title={Beyond the standard quantum limit for parametric amplification of broadband signals},
journal={npj Quantum Inf.},
year={2021},
month={Nov},
day={08},
volume={7},
number={1},
pages={160},
issn={2056-6387},
doi={10.1038/s41534-021-00495-y},
url={https://doi.org/10.1038/s41534-021-00495-y}
}

\end{document}